\documentclass[showpacs,aps,prb,preprint,floatfix,superscriptaddress]{revtex4}
\begin{document}
\title{Exchange-correlation energy densities for two-dimensional
  systems from quantum dot ground-states}
\author{Andreas Wensauer}
\affiliation{Institut f\"ur Theoretische Physik, 
         Universit\"at Regensburg, D-93040 Regensburg, Germany}
\author{Ulrich R\"ossler}
\affiliation{Institut f\"ur Theoretische Physik, 
         Universit\"at Regensburg, D-93040 Regensburg, Germany}
\date{\today}

\begin{abstract}

In this paper we present a new approach how to extract
polarization-dependent exchange-correlation energy densities for
two-dimensional 
systems from reference densities and energies of quantum 
dots provided by exact diagonalization. Compared with results from
literature we find systematic 
corrections for all polarizations in the regime of high densities.

\end{abstract}

\pacs{73.21.-b, 71.10.Ca, 71.15.Mb}
 
\maketitle

\section{Introduction}

Density functional theory (DFT)\cite{hohenberg64, kohn65} and
spin-density functional theory (SDFT)\cite{barth72, rajagopal73}
are powerful techniques to investigate interacting electron
systems. However, the results of these methods sensitively depend on
the quality of 
the approximation of the exchange-correlation (XC) functional by the
local (spin) density approximation (L(S)DA) and/or gradient corrections.
In contrast to three-dimensional (3D) systems where a large number
of parameterizations 
for XC energy densities and gradient corrections is available the
situation is different for two-dimensional (2D) systems. The majority of
all calculations for 2D (e.g.\ for QDs see Refs.\
[\onlinecite{ortiz93, koskinen97, austing99, hirose99, steffens99,
  wensauer00, wensauer01, ciorga02, wensauer03}]) relies on
the parameterization of the XC energy density from Tanatar and
Ceperley\cite{tanatar89} (TC). They numerically calculated XC energy
densities as a function of the carrier density (density parameter
$r_s$) and spin polarization $\xi$ by applying variational Monte
Carlo and fixed-node Green's 
function Monte Carlo methods. The results for $\xi=0$ (no
polarization) and $\xi= 1$ 
(full polarization) and  $r_s$-values 1, 5, 10, 15, 20 etc.\ were
fitted and interpolated using a Pad\'e approximation
\begin{equation}
\varepsilon_{\rm{c}}^{\rm{TC}}(r_s,\xi)/\mbox{Ry}
=a_0 {1+ a_1 x\over 1 + a_1 x + a_2 x^2 + a_3 x^3} \quad\mbox{($x=r_s^{1/2}$)}.
\label{TC}
\end{equation}
with parameters $a_j$ for $\xi=0$ and $ 1$ respectively (see
Tab.\ IV in Ref.\ [\onlinecite{tanatar89}]). In the limit of
$r_s\rightarrow 0$ the 
asymptotic behavior is characterized by $a_0\left[1+b
  r_s+O\left(r_s^{3/2}\right)\right]$ whereby logarithmic
contributions (see ring-diagram summation in Ref.\
[\onlinecite{rajagopal77}]) such as
$r_s\ln r_s$ are not taken into account. The construction of the
dependence on the polarization $\xi$ is analogous to the 3D
case\cite{barth72, rajagopal73}
where the interpolation of the exchange (X) energy in
Hartree-Fock approximation  
$f(\xi)=\left[(1+\xi)^{3/2}+(1-\xi)^{3/2}-2^{3/2}\right]/(2-2^{3/2})$
is used
\begin{eqnarray}
&\varepsilon_{\rm{XC}}^{\rm{TC}}(r_s,\xi)/\mbox{Ry}
=&-{4\sqrt{2}\over 3\pi
  r_s}\left[(1+\xi)^{3/2}+(1-\xi)^{3/2}\right]\nonumber\\ 
& &+\varepsilon_{\rm{c}}^{\rm{TC}}(r_s,\xi=1)/\mbox{Ry}
+\left[\varepsilon_{\rm{c}}^{\rm{TC}}(r_s,\xi=0)
-\varepsilon_{\rm{c}}^{\rm{TC}}(r_s,\xi=1)\right]f(\xi)/\mbox{Ry}.
\label{TCformel}
\end{eqnarray}

%

Attaccalite, Moroni, Gori-Giorgi, and Bachelet\cite{attaccalite02}
(AMGB) pursue a 
similar approach. They calculate the ground state (GS) energy of a 2DEG as a
function of $r_s$ and $\xi$ with fixed-node diffusion Monte Carlo
methods which also take into account back-flow correlations. Their
parameterization of the XC energy density
\begin{equation}
\varepsilon_{\rm XC}^{\rm AMGB}(r_s,\xi)/{\rm Ry}=\varepsilon_{\rm
  X}(r_s,\xi)/{\rm Ry}+\varepsilon_{\rm C}(r_s,\xi)/{\rm Ry} 
\label{ambg}
\end{equation}
comprises the well-known X energy from the ring-diagram method\cite{rajagopal77}  
\begin{equation}
\varepsilon_{\rm X}(r_s,\xi)/{\rm Ry}=
-{4\sqrt{2}\over 3\pi r_s}\left[(1+\xi)^{3/2} + (1-\xi)^{3/2}\right]
\end{equation}
and the correlation energy
\begin{equation}
\varepsilon_{\rm C}(r_s,\xi)/{\rm Ry}=
\left({\rm e}^{-\beta r_s} -1\right)\varepsilon_{\rm
  X}^{(6)}(r_s,\xi)/{\rm
  Ry}+2\alpha_0(r_s)+2\alpha_1(r_s)\xi^2+2\alpha_2(r_s)\xi^4{\text ,}
\label{ambg1}
\end{equation}
with 
\begin{equation}
\varepsilon_{\rm X}^{(6)}(r_s,\xi)=\varepsilon_{\rm
  X}(r_s,\xi)-\left(1+\frac{3}{8} \xi^2+\frac{3}{128}\xi^4\right)
\varepsilon_{\rm X}(r_s,0)
\end{equation}
being the Taylor expansion of the X energy with respect to $\xi$ of
the order six and higher. This kind of representation of the XC
energy allows for the identification of the term $2\alpha_0(r_s)$
with the correlation energy $\varepsilon_{\rm C}(r_s,0)$ and of the
term $2\alpha_1(r_s)$ with the spin stiffness. The
$\alpha_i(r_s)$-parameterizations are generalizations of the
Perdew-Wang form\cite{perdew92} to 2D
\begin{equation}
\alpha_i(r_s)=A_i+\left(B_ir_s+C_ir_s^2+D_ir_s^3
\right)\log\left(1+\frac{1}{E_ir_s+F_ir_s^{3/2}+G_ir_s^2+H_ir_s^3}\right). 
\label{ambg2}
\end{equation}
The parameters which were fitted to the numerical results are
summarized in Tab.\ II of Ref.\ [\onlinecite{attaccalite02}]. 

%

Another recent work on 2D XC energy densities\cite{seidl01} is based 
on a purely analytical approach. The
interaction-strength-interpolation (ISI) connects the two limits of
strongly and weakly interacting 2DEGs. The result for the XC energy
density reads 
\begin{eqnarray}
\varepsilon_{\rm XC}(r_s,\xi)/{\rm Ry}&=&
2\frac{a_\infty}{r_s}+4\frac{X(r_s,\xi)}{Y(r_s,\xi)}\times
\nonumber\\
&&\times\left[(1+Y(r_s,\xi))^{1/2}-1-Z(\xi)\log{\frac{(1+Y(r_s,\xi))^{1/2}+Z(\xi)}
{1+Z(\xi)}}\right].
\label{ISI}
\end{eqnarray}
The functions and parameters entering the interpolation are 
\begin{eqnarray}
&&X(r_s,\xi)=\frac{-b_0(\xi)}{r_s\left[c_x(\xi)-a_\infty\right]^2}\text{,}\\
&&Y(r_s,\xi)=\frac{4b_0(\xi)^2r_s}{\left[c_x(\xi)-a_\infty\right]^4}\text{,}\\
&&Z(\xi)=\frac{-b_0(\xi)}{\left[c_x(\xi)-a_\infty\right]^3}-1
\end{eqnarray}
and
\begin{eqnarray}
&&a_\infty=-(2-8/(3\pi))\text{,}\\
&&b_0(\xi)=0.1125\xi^8-0.1495\xi^6+0.083\xi^4+0.107\xi^2-0.192\text{,}\\
&&c_x(\xi)=-{2\sqrt{2}\over 3\pi}\left[(1+\xi)^{3/2} + (1-\xi)^{3/2}\right].
\end{eqnarray}

Two observations can be made with respect to these different forms
of the XC energy densities:
\begin{enumerate}
\item{To visualize the different parameterizations we show in
    Fig.~\ref{fig_exccomp} the relative deviations, i.e.\
    $\varepsilon_{\rm XC}^{(1)}/\varepsilon_{\rm 
  XC}^{(2)}-1$. Figs.~\ref{fig_exccomp} (a) and (b) compare the
AMGB- and TC-XC energy density on different scales for $r_s$. We
find a good agreement 
  ($\varepsilon_{\rm XC}^{\rm AMGB}/\varepsilon_{\rm 
  XC}^{\rm TC}-1\le 0.008$) which is not surprising as both expressions
are based on the same method. On the other hand it is nevertheless
remarkable because AMGB explicitly calculated polarizations
$0<\xi<1$ whereas TC interpolated over this regime. Larger
deviations can be seen for $\varepsilon_{\rm
  XC}^{\rm ISI}/\varepsilon_{\rm XC}^{\rm TC}-1$ in (c) and for
$\varepsilon_{\rm 
  XC}^{\rm ISI}/\varepsilon_{\rm XC}^{\rm AMGB}-1$ in (d). There, the
relative differences go up to $0.04$, especially for full
polarization. This deviation should have its origin in the different
applied methods (ISI analytical, TC and AMGB numerical).}  
\item{Besides looking at the difference between the various XC
    energy densities we 
can also examine their performance when they are used as input in DFT.
In a recent paper\cite{reimann00} the agreement between
SDFT/LSDA (using the TC parameterization) and ED densities for QDs was
investigated and found to be fairly 
good. However this has not to be true  
for all systems. For example in Fig.~\ref{fig_motivation} we compare GS
densities of a QD with $N=8$, $\omega=3{\rm meV}$,
$S=S_z=4$, $L=0$ ($\hbar=1$ in this paper). Especially for the
density maximum and minimum the 
difference between ED and SDFT/LSDA is considerably large. Further,
this effect does not depend on the applied parameterization.}
\end{enumerate}

These two findings are our motivation to check the accuracy of XC
energy densities in 2D by following a
new approach. Our starting point are GS densities and energies from
ED from which the XC energy densities are extracted in two steps:
First we design an iterative scheme to calculate the self-consistent XC
potentials of the KS 
equations which reproduce the exact densities. The second step deals
with the extraction of XC energy 
densities from these XC potentials assuming L(S)DA.

Our paper is organized as follows: We briefly introduce
the QD Hamiltonian in Sec.\ II and the SDFT/LSDA in Sec.\ III. Then
we describe how to extract XC potentials from GS densities (Sec.\
IV) and XC energy densities from XC potentials (Sec.\ V). Sec.\ VI
contains the results for systems with arbitrary polarization.
We end with a short summary of the most important conclusions in
Sec.\ VII.  

\section{Quantum dot Hamiltonian and ground state densities}

We consider a two-dimensional QD with an axially symmetric
parabolic confinement potential of strength $\omega_0$.
As we are especially interested in GS configurations with vanishing
angular momentum $L=0$ a
magnetic field can be omitted. 
The Hamiltonian for $N$ particles in real-space representation [with
${\bf r}=(x,y)$, ${\bf p}= (p_x,p_y)$] reads: 
\begin{eqnarray}
\hat H=\sum_{j=1}^N \left(\frac{{\bf p}_j^2}{2m^\ast} +
\frac{1}{2}m^\ast\omega_0^2{\bf r}_j^2 \right)
+\frac{1}{2} \sum_{j,k=1}^{N}{\!\! ^\prime}\,\,
\frac{e^2}{4\pi\varepsilon\varepsilon_0|{\bf r}_j-{\bf r}_k|}.
\label{ham_rss}
\end{eqnarray}
Here $m^\ast$ is the effective mass, $e$ is the electron charge,
and $\varepsilon$ is the screening constant of the host semiconductor.

In order to get the reference densities and energies for the
investigation of XC energy densities we apply ED techniques which
provide results of high accuracy.\cite{wensauer03b} 
Please note that the (spin-)density of all eigenstates of the angular
momentum operator are functions of radius $r$ but not of the angle
$\varphi$. Therefore, the relevant quantities from ED are the
spin-densities $n_\sigma(r)$ of the GS and its energy $E_0$.  

\section{Spin-density functional theory and local (spin-)density
  approximation} 

In this Section we sketch the basics of (S)DFT and L(S)DA. 
The DFT formalism was originally established by
Hohenberg, Kohn, and Sham\cite{hohenberg64, kohn65} and generalized to 
spin-polarized systems\cite{barth72} by including the
coupling of the polarization to an applied magnetic field. Accordingly,
the Hohenberg-Kohn (HK) theorem has to be modified with respect
to the spin degrees of freedom\cite{barth72}. For this case, it states that two
different non-degenerate ground-state wavefunctions $|\Psi\rangle$ and
$|\Psi^\prime\rangle$ always yield different combinations $(n_\sigma({\bf
  r}))\neq (n^\prime_\sigma({\bf r}))$ of 
spin densities. This is sufficient to establish a functional of the total
energy with the usual functional properties
\begin{equation}
E_{V_{\sigma}}[n_\sigma]=F_{\rm HK}[n_\sigma]+\sum_\sigma\int{\rm d}{\bf
  r}V_\sigma({\bf r})n_\sigma({\bf r})
\end{equation}
and the universal HK functional 
\begin{equation}
F_{\rm HK}[n_\sigma]=\langle\Psi[n_\sigma]|T+W|\Psi[n_\sigma]\rangle .
\end{equation}
Thus, even in the limit of vanishing magnetic fields the SDFT scheme can
yield a spin-polarized
ground state for even electron numbers due to Hund's 
rule.\cite{koskinen97, steffens98a}

For practical purposes the variational scheme has to be mapped on
the Kohn-Sham (KS) system, i.e.\ an effective single-particle system with
the same GS densities as the interacting system.
The spin-degree of freedom is considered in the KS equations\cite{barth72} by
assuming the total spin $S_z$ in $z$-direction to be a good quantum number
\begin{equation}
\left\lbrace -\frac{\hbar^2}{2m^\ast}\nabla^2+V_\sigma({\bf r})+
\frac{e^2}{4\pi\varepsilon\varepsilon_0} \int{\rm d}{\bf r}^\prime
\frac{n({\bf r}^\prime)}{\left|{\bf r}-{\bf r}^\prime\right|}+
V_{{\rm XC},\sigma}([n_\sigma],{\bf r})\right\rbrace\varphi_{j,\sigma}({\bf r})
=\varepsilon_{j,\sigma}\varphi_{j,\sigma}({\bf r})
\end{equation}
with the spin $\sigma=\pm$ in $z$-direction and the KS energies
$\varepsilon_{1,\sigma}\le\varepsilon_{2,\sigma}\le...$ . For a
system containing $N$ particles 
the spin densities are given by
\begin{equation}
n_\sigma({\bf r})=
\sum_j\gamma_{j,\sigma}\left|\varphi_{j,\sigma}({\bf r})\right|^2
\end{equation}
with $\gamma_{j,\sigma}$ being occupation numbers
of the KS levels in the ground state 
($\sum_j \gamma_{j,\sigma}=N_\sigma$ and $N_++N_-=N$).
Then the GS density and polarization are
\begin{equation}
n({\bf r})=n_+({\bf r})+n_-({\bf r})
\end{equation}
\begin{equation}
\xi({\bf r})=\frac{n_+({\bf r})-n_-({\bf r})}{n({\bf r})}.
\end{equation}
The XC potentials 
\begin{equation}
V_{\rm XC,\sigma}([n_\sigma],{\bf r})=
\frac{\delta E_{\rm XC}[n_\sigma]}{\delta n_\sigma({\bf r})}
\end{equation}
are defined as functional
derivatives of the XC energy functional
\begin{equation}
E_{\rm XC}[n_\sigma]=F_{\rm HK}[n_\sigma]-
\frac{1}{2}\frac{e^2}{4\pi\varepsilon\varepsilon_0}
\int{\rm d}{\bf r}\int{\rm d}{\bf r}^\prime
\frac{n({\bf r})n({\bf r}^\prime)}
{\left|{\bf r}-{\bf r}^\prime\right|}-T_{\rm S}[n_\sigma].
\label{vxcdef}
\end{equation}
($T_{\rm S}[n_\sigma]$ denotes the kinetic energy functional of the KS
system.)
The total ground-state energy $E_0$ of the interacting system can be calculated
from 
\begin{equation}
E_0=\sum_{j,\sigma}\gamma_{j,\sigma}\varepsilon_{j,\sigma}-
\frac{1}{2}\frac{e^2}{4\pi\varepsilon\varepsilon_0}
\int{\rm d}{\bf r}\int{\rm d}{\bf r}^\prime
\frac{n({\bf r})n({\bf r}^\prime)}
{\left|{\bf r}-{\bf r}^\prime\right|}
-\sum_\sigma\int{\rm d}{\bf r}
\,V_{\rm XC,\sigma}([n_\sigma],{\bf r})n_\sigma({\bf r})+E_{\rm
  XC}[n_\sigma]. 
\label{Eformel}
\end{equation}
Concerning the XC potentials we apply the L(S)DA 
\begin{equation}
E_{\rm XC}[n_\sigma]\approx
\int{\rm d}{\bf r}\,n({\bf r})\,
\varepsilon_{\rm XC}(n_+({\bf r}),n_-({\bf r})).
\label{lvsdaeq}
\end{equation}
The most important parameterizations for the XC energy density
$\varepsilon_{\rm XC}(n_+,n_-)$ (or $\varepsilon_{\rm XC}(r_s,\xi)$)
used in 2D calculations were introduced in Sec.\ I. 

\section{Calculation of XC potentials}
\label{n2v}

The first step of our concept to obtain XC energy densities for 2D
systems is the calculation of XC potentials from GS densities.  
Before presenting our method we will briefly review the literature
on the inversion of the KS equations for 3D systems. 
In a pioneer work Almbladh and Pedroza adapt parametrized XC
potentials to the electron densities of light atoms.\cite{almbladh84} 
An alternative approach by Aryasetiawan and Stott formulates
the problem in terms of $(N-1)$ coupled non-linear differential 
equations.\cite{aryasetiawan88} Holas and March derive a solution by
applying the Pauli potential and energy and $(N-1)$ Euler equations
leading to a differential equation (DEQ) for the density amplitude
$\sqrt{n({\bf r})}$.\cite{holas91} G\"orling describes an approach
based on the linear response of potentials on small density
modifications.\cite{goerling92} An iterative method to construct KS
orbitals and XC potentials for a given electron density is presented
by Wang and Parr\cite{wang93}. They use the inverted
KS-Schr\"odinger equation to generate an improved effective
potential for the next iteration. In another paper which mainly
focuses on kinetic energy functionals Zhao et al.\ propose a method
which is based on Lagrange multipliers to gain XC potentials.\cite{zhao94}
Tozer et al.\ train neural networks\cite{tozer96} using XC
potentials calculated with Zhao et al.'s method and determine
fit parameters for XC functionals\cite{tozer97}.   
Following Ref.\ [\onlinecite{wang93}] Leeuwen and
Baerends present a modified form of the inverted KS-Schr\"odinger
equation\cite{leeuwen94} which will also be applied for a 2D system
in the present paper and be discussed in detail below.

The KS-Hamiltonian of axially symmetric 2D QDs reads
\begin{equation}
H_{{\rm S},\sigma}=-\frac{\hbar^2}{2m^\ast}\left\lbrack
  \frac{1}{r}\frac{\partial}{\partial
  r}\left(r\frac{\partial}{\partial r} \right)+\frac{1}{r^2}\frac{\partial^2}{\partial\varphi^2} \right\rbrack
  +\frac{1}{2}m^\ast\omega_0^2r^2+V_{\rm H}(r)+V_{\rm
  XC,\sigma}([n_\sigma],r) 
\label{effksham}
\end{equation}
Inversion of the KS equations means the calculation of the exact
XC potentials up to a gauge-constant $c_\sigma$
\begin{equation}
V_{{\rm XC},\sigma}(r)
:=V_{{\rm XC},\sigma}([n_\sigma],r)+c_\sigma
\label{vxc}
\end{equation}
of a system with given electron spin densities $n_\sigma(r)$, i.e.\ we want to
find the self-consistent solution of the KS-Schr\"odinger equation
(\ref{effksham}) under the constraint
$n_\sigma(r)=n_{{\rm KS},\sigma}(r)$.
The iteration scheme\cite{leeuwen94} is based on a method
presented in Ref.\ [\onlinecite{wang93}]. It should not be considered as
a strict proof but as plausibility argument.

In a first step we split off the scalar XC potential from the KS-Hamiltonian (\ref{effksham})
and denote the rest by $\hat H_{0,\sigma}$ 
\begin{equation}
\hat H_{{\rm S},\sigma}=\hat H_{0,\sigma}+\hat V_{{\rm XC},\sigma}.
\end{equation}
Using the eigenfunctions $\varphi_{j,\sigma}({\bf r})$ and the eigenvalues
$\varepsilon_{j,\sigma}$  of the KS-Hamiltonian we obtain the link
between exact spin densities and exact XC-potentials
\begin{equation}
V_{{\rm XC},\sigma}(r)=\frac{1}{n_\sigma(r)}\sum_j\gamma_{j,\sigma}
\varphi_{j,\sigma}^\ast({\bf r})
\left(\varepsilon_{j,\sigma}-H_{0,\sigma}\right)\varphi_{j,\sigma}({\bf r}).
\label{magic}
\end{equation} 
We denote eigenfunctions, eigenvalues, and densities of the $k$-th
iteration step by $\varphi_{j,\sigma}^{(k)}({\bf r})$,
$\varepsilon_{j,\sigma}^{(k)}$, and $n_{{\rm
KS},\sigma}^{(k)}(r)$, i.e.\ they are solutions of the
KS-Schr\"odinger equation with the KS-Hamiltonian $\hat H_{{\rm
    S},\sigma}^{(k-1)}=\hat H_{0,\sigma}+\hat V_{{\rm
    XC},\sigma}^{(k-1)}$. 
Thus, we are able to construct the XC potential for the next step
by applying Eq.~(\ref{magic})
\begin{equation}
V_{{\rm XC},\sigma}^{(k)}(r)=\frac{1}{n_\sigma(r)}\sum_j\gamma_{j,\sigma}
{\varphi_{j,\sigma}^{(k)}}^\ast({\bf r})
\left(\varepsilon_{j,\sigma}^{(k)}-H_{0,\sigma}\right)\varphi_{j,\sigma}({\bf r})^{(k)}.
\label{magic2}
\end{equation} 
With wavefunctions $\varphi_{j,\sigma}^{(k)}({\bf r})$ being solutions of $\hat H_{{\rm
S},\sigma}^{(k-1)}$, the iteration scheme for scalar
potentials\cite{leeuwen94} is given by 
\begin{equation}
V_{{\rm XC},\sigma}^{(k)}(r)=
V_{{\rm XC},\sigma}^{(k-1)}(r)\frac{n_{{\rm
KS},\sigma}^{(k-1)}(r)}{n_\sigma(r)}.
\label{vxcadapt}
\end{equation}
During the iteration we will assume $V_{{\rm XC},\sigma}^{(k)}>0$. 
If the KS density in the $(k-1)$-th iteration step is locally too
high (low) the potential of step $k$ will be reduced
(increased) at the same place. Consequently, the new density will be
larger (smaller). However, the gauge constants $c_\sigma$ of
the potentials $V_{{\rm XC},\sigma}^{(k)}$ cannot be determined by
this method (see Ref.\ [\onlinecite{capelle01}]). We will discuss this
problem in detail in the following Section.  


In the context of the iteration process the aspect of
representability of the exact GS density is also tested: As result
of a converging iteration we obtain an effective potential so that
the KS density reproduces the density of the exact system. 

We tested the numerical procedure for a QD with six electrons and a
confinement potential of $3{\rm meV}$. In the GS the two lowest
shells are occupied, i.e.\ the system is unpolarized and the
paramagnetic current density vanishes due to $L=0$. In this test
calculation we started from a GS density which was calculated
using conventional DFT/LDA and TC parameterization for XC
energies. Thus, we could exclude any problems arising from
representability. The result of this test calculation shows perfect
agreement between initial and final densities and effective
potentials proving the validity of our program.

\section{Calculation of XC energy densities}
\label{v2e}

In this Section we focus on the second step of the system, the
extraction of XC energy densities 
from exact XC potentials. 
In order to establish a relation between these two quantities we
apply the L(S)DA (\ref{lvsdaeq}) on the general Eq.\ (\ref{vxcdef})
for the functionals 
and obtain
\begin{equation}
V_{{\rm XC},\sigma}(r)=\varepsilon_{\rm XC}(r)
+n(r)\frac{\partial \varepsilon_{\rm XC}}{\partial n_\sigma}(r).
\end{equation}
We solve for $\frac{\partial \varepsilon_{\rm XC}}{\partial
  n_\sigma}(r)$ 
\begin{equation}
\frac{\partial \varepsilon_{\rm XC}}{\partial n_\sigma}(r)
=\frac{1}{n(r)}\left(-\varepsilon_{\rm XC}(r)+V_{{\rm
      XC},\sigma}(r)\right) 
\label{vxc1}
\end{equation}
and plug the result into the derivative of the XC energy density as
a function of the radius
\begin{equation}
\frac{\partial \varepsilon_{\rm XC}(r)}{\partial r}=
\sum_\sigma \frac{\partial \varepsilon_{\rm XC}}{\partial n_\sigma}
\frac{\partial n_\sigma}{\partial r}.
\end{equation}
After some algebra we arrive at a linear DEQ
\begin{equation}
\frac{\partial \varepsilon_{\rm XC}(r)}{\partial r}+
\frac{\partial\log n}{\partial r}(r)\varepsilon_{\rm XC}(r)=I(r)
\label{deq}
\end{equation}
with the inhomogeneity
\begin{eqnarray}
I(r)=&\frac{1}{n(r)}\sum_\sigma \left(V_{{\rm
XC},\sigma}(r)+c_\sigma\right) 
\frac{\partial n_\sigma}{\partial r}(r)
\label{inh}
\end{eqnarray}
which contains all the information about the dependence on the spin
densities. 
In Eq.\ (\ref{inh}) we take into account that the gauge constants
$c_\sigma$ of the scalar potentials are not known from the previous
step. They will be calculated later.
The solution of the homogeneous part of DEQ (\ref{deq})
is given by
$\varepsilon_{\rm XC}^{\rm hom}(r)=\alpha/n(r)$,
a special solution can be calculated using the ansatz
$\varepsilon_{\rm XC}^{\rm spez}(r)=\beta(r)/n(r)$.
The function $\beta(r)$ results from an elementary DEQ
\begin{equation}
\frac{\partial \beta}{\partial r}(r)=n(r)I(r){\text ,}
\end{equation}
whose solution is
\begin{equation}
\beta (r)=\int\limits_0^r{\rm
d}r^\prime\,n(r^\prime)I(r^\prime)-\beta(0).
\end{equation}
Thus the general solution of DEQ (\ref{deq}) is
\begin{equation}
\varepsilon_{\rm XC}(r)=\varepsilon_{\rm XC}^{\rm
hom}(r)+\varepsilon_{\rm XC}^{\rm
spez}(r)=\frac{\alpha-\beta(0)}{n(r)} 
+\frac{1}{n(r)}\int\limits_0^r{\rm
d}r^\prime\,n(r^\prime)I(r^\prime).
\label{gen}
\end{equation}
After substituting the inhomogeneity the result reads
\begin{eqnarray}
\varepsilon_{\rm XC}(r)=&\frac{\alpha-\beta(0)
-n_\sigma(0)\sum_\sigma c_\sigma}{n(r)}+\sum_\sigma c_\sigma\frac{n_\sigma(r)}{n(r)}
+\frac{1}{n(r)}
\int\limits_0^r{\rm d}r^\prime
\sum_\sigma V_{{\rm XC},\sigma}(r^\prime) 
\frac{\partial n_\sigma}{\partial r^\prime}(r^\prime).
\end{eqnarray}
With the modulus of the XC energy being finite in L(S)DA
\begin{equation}
|E_{\rm XC}[n_\sigma]|\approx \left|\int {\rm d}{\bf r}
 \,n(r)\varepsilon_{\rm XC}(r)\right|= \int {\rm d}{\bf r}
 \,n(r)|\varepsilon_{\rm XC}(r)| <\infty
\end{equation}
we choose $\alpha-\beta(0)-n_\sigma(0)\sum_\sigma c_\sigma=0$ thus
avoiding any divergent contributions.
Consequently the analytical solution of (\ref{deq}) satisfying the
physical boundary conditions is
\begin{eqnarray}
\varepsilon_{\rm XC}(r)=\sum_\sigma c_\sigma\frac{n_\sigma(r)}{n(r)}
+ \frac{1}{n(r)}
\int\limits_0^r{\rm d}r^\prime
\sum_\sigma & V_{{\rm XC},\sigma}(r^\prime) 
\frac{\partial n_\sigma}{\partial r^\prime}(r^\prime).
\label{exc2}
\end{eqnarray}

The last step is the calculation of the gauge constants $c_\sigma$
of the scalar potentials. One condition which has not been used
yet is the agreement of the DFT GS energy and the exact GS energy.
If we eliminate $V_{{\rm XC,}\sigma}({\bf r})$ in Eq.\ (\ref{Eformel})
and write it in a modified form
\begin{equation}
E_{GZ}=\sum_{j\sigma}\gamma_{j\sigma}
\int {\rm d}{\bf r}\,\varphi_{j\sigma}^\ast({\bf r})
\left(\frac{{\bf p}^2}{2m^\ast}+V_\sigma({\bf r})\right)
\varphi_{j\sigma}({\bf r})+\frac{1}{2}\int {\rm d}{\bf r}\,V_{\rm H}({\bf r})n({\bf r})
+E_{\rm XC}[n_\sigma]
\label{ksgs}
\end{equation}
we can calculate the (exact!) XC energy $E_{\rm XC}[n_\sigma]$ for
the GS densities $(n_\sigma)$: the KS-wavefunctions are known from
the selfconsistent 
solution of the KS equations (see \ref{n2v}). Thus, the expectation
values $\int {\rm d}{\bf r}\,\varphi_{j\sigma}^\ast({\bf r})
\left(\frac{1}{2m^\ast}{\bf p}^2+V_\sigma({\bf r})\right) 
\varphi_{j\sigma}({\bf r})$ of the non-interacting system can be
calculated. The Coulomb energy and the GS-energy are uniquely
determined by the interacting system.
On the other hand the XC energy for QDs in L(S)DA is given by
\begin{equation}
E_{\rm XC}[n_\sigma]\approx
\int{\rm d}{\bf r}\,n(r)
\varepsilon_{\rm XC}(r)
=\sum_\sigma c_\sigma N_\sigma+\int\limits_{0}^{\infty}{\rm d}r
\int\limits_0^r{\rm d}r^\prime
\sum_\sigma V_{{\rm XC},\sigma}(r^\prime)
\frac{\partial n_\sigma}{\partial r^\prime}(r^\prime)
{\text ,}
\label{ksgs2}
\end{equation}
what makes $\sum_\sigma c_\sigma N_\sigma$ accessible.
In the case of unpolarized systems $c:=c_\uparrow=c_\downarrow$ 
and for full polarization $c_\sigma$ of the unoccupied spin
direction is irrelevant (as the corresponding $N_\sigma=0$). Then the results for the XC
energy density are unique. For partially  
polarized systems uniqueness of the results can be achieved by
additionally demanding asymptotic agreement of
$V_{\rm XC,\sigma}(r)$ for finite systems\cite{capelle01}  
\begin{equation}
\lim\limits_{r\to\infty}V_{\rm
  XC,\uparrow}(r)=\lim\limits_{r\to\infty}V_{\rm XC,\downarrow}(r). 
\end{equation}

As a result of the inversion of the L(S)DA formalism we obtain the
XC energy density as a function of the radius. After eliminating the
radius by using the spin densities $n_\sigma(r)$, or alternatively $r_s(r)$
and $\xi(r)$, we arrive at the standard representation of the XC
energy density as a function of $(r_s,\xi)$. 

\section{Results}

After introducing the methodology in the two previous Sections we
summarize the numerical results for XC potentials 
and XC energy densities here. The investigated systems are classified due to
their degree of polarization. For unpolarized ($\xi=0$) and fully
polarized systems ($\xi=1$) the XC energy density is only a function
of the density parameter $r_s$ whereas for partially polarized
systems $\varepsilon_{\rm XC}$ depends on both $r_s$ and
polarization $\xi$. But first we will examine an analytically
solvable system for two electrons to study the asymptotics of XC
potentials. 

\subsection{The two-electron system}

The system with two electrons will be used to find the
asymptotic behavior of densities, KS wavefunctions, and XC
potentials as QD helium can be calculated analytically for special values
of the confinement potential.\cite{taut94}
For a system with confinement energy
$\omega=2\,{\rm Ry}$ the analytical expression for the singlet GS
wavefunction (GS energy $6\,{\rm Ry}$) reads
\begin{eqnarray}
\Psi({\bf r}_1,{\bf r}_2)&=&\varphi_{\rm CM}(|{\bf r}_1+{\bf
r}_2|/2)\,\varphi_{\rm rel}(|{\bf r}_1-{\bf r}_2|)\nonumber\\
&=& C_{\rm CM} {\rm e}^{-|{\bf r}_1+{\bf r}_2|^2/4}\cdot C_{\rm
  rel}(1+r){\rm e}^{-|{\bf r}_1-{\bf r}_2|^2/4} 
\end{eqnarray} 
with $C_{\rm CM}=\sqrt{2/\pi}$ and $C_{\rm
  rel}=1/\sqrt{2\pi(\sqrt{2\pi}+3)}$.\cite{taut94}
Consequently, we are also able to
  calculate the density, the Hartree potential and the XC potential
  analytically. As the full expressions
  of these quantities are very complex and do not give much insight
  we restrict ourselves to the formulas which yield the correct
  asymptotics for the limit of large radii. The density decays
  exponentially with
\begin{eqnarray}
\lim\limits_{r\to\infty}n({r})=2\pi C_{\rm CM}^2C_{\rm rel}^2 r^2
{\rm e}^{-r^2} 
\end{eqnarray}
and the Hartree potential for large $r$ is that of a point charge
\begin{eqnarray}
\lim\limits_{r\to\infty}V_{\rm H}({r})=4/r.
\end{eqnarray}
Thus, the asymptotics of the KS wavefunction $R({r})=\sqrt{\pi
  n({r})}$ is 
\begin{eqnarray}
\lim\limits_{r\to\infty}R({r})=\sqrt{2}\pi C_{\rm CM}C_{\rm rel}
r {\rm e}^{-r^2/2}. 
\end{eqnarray}
$R({r})=\sqrt{\pi n({r})}$ can be used to calculate the XC
potential
\begin{eqnarray}
V_{\rm XC}({r})/{\rm Ry}=\varepsilon_0-r^2-V_{\rm
  H}({r})+\frac{\left\lbrack\frac{1}{r}\frac{\partial}{\partial 
  r}(r\frac{\partial}{\partial r}) \right\rbrack R({r})}{R({r})}.
\label{2elxc}
\end{eqnarray}
($r^2$ is the parabolic confinement in ${\rm Ry}$.)
The KS energy $\varepsilon_0=4\,{\rm Ry}$ was chosen so that the the
XC potential vanishes in the limit $r\to \infty$
\begin{eqnarray}
\lim\limits_{r\to\infty}V_{\rm XC}({r})=-2/r+1/r^2.
\end{eqnarray}
The leading term $-2/r$ is a manifestation of the strong
selfinteraction in the case of two electrons. The selfinteraction
potential is exactly compensated by the X potential $V_{\rm
  X}(r)=V_{\rm H}(r)/2$ in the case of a singlet
configuration.\cite{filippi94} 

In summary we found from this analytical calculation that the
asymptotic behavior of the XC potential is a rational function which
converges to zero. XC energy densities cannot be extracted from this
kind of system because of the dominating selfinteraction
contribution.  

\subsection{Unpolarized systems}

Unpolarized configurations without angular
momentum are typically closed-shell systems with 2, 6, 12, etc.\
electrons. Unfortunately, the GS densities from ED for 12 and more particles
are not convergent so far, so that they cannot be used as reference
densities at present. On the other hand, ED will give excellent results for two
particles. However, predominating self-interaction effects in a
two-electron system are a problem for the extraction of reliable XC
energy densities. Therefore, we will focus on the six-electron
system with two closed shells.

To illustrate the inversion of the KS
equations we investigate a six-electron system with a standard
confinement potential 
$3.32\,{\rm meV}$ (see Fig.~\ref{fig_unpolexample}). The reference
GS density from ED is depicted in 
(a). The quality of the ED results was tested using an increasing
number of Slater determinants. For the present calculation with
326120 Slater determinants we found a good convergence of GS
densities up to a radius of $8\,{\rm a}_0$. 

By applying the iteration scheme (\ref{vxcadapt}) we are now able to calculate
the XC potential up to a constant (see black dotted line in (b)) which exactly
reproduces the reference density from ED. For reasons of comparison
there is also the TC XC-energy density plotted in (b) which was
calculated with the reference density. In the next step we solve the
DEQ (\ref{deq}) to calculate the XC energy density as a function of
the radius and determine the gauge constant $c_\sigma$. The
extracted XC energy density (black dashed line) is 
shown together with the TC XC energy density (grey solid) in (c). In (d)
the same two curves are plotted, however,
we eliminated the radius in favor of the density parameter. The
redundancy in the extracted XC energy density (black, dashed curve)
is explained as follows: Close to the local density maximum we get two
$\varepsilon_{\rm XC}$ values per density value as
reflected in Fig.~\ref{fig_unpolexample} (d). The deviation between
the two values will be discussed later. 

First let us focus on the difference in the high-density region in the
center and in the low-density regime at the edge (see
Fig.~\ref{fig_unpolexample}). While the
results for high densities are reliable, the differences at the edge
of the dot are a consequence of the finite basis set in ED. 
However, in the analytically solvable model for two
electrons the XC potential ({\ref{2elxc}) converges to
zero, whereas this is not true for the ED result (see
Fig.~\ref{fig_unpolexample}(b)). Consequently, the problem emerges again in the
XC energy density (see Fig.~\ref{fig_unpolexample}(c),
(d)). Thus, we can reassure that the behavior of $V_{\rm XC}(r)$ at
the edge is not due to the divergence of $(\partial n(r)/\partial
r)/n(r)$ or non-local effects.
However, we want to emphasize that the difficulties of XC
energy densities at the edge do not affect the accuracy of the gauge
constant $c_\sigma$ as in Eq.~(\ref{ksgs}) $\varepsilon_{\rm
  XC}(r)$ is weighted with the exponentially decreasing density.

Now let us turn to the redundant structure in $\varepsilon_{\rm XC}$ of
Fig.~\ref{fig_unpolexample} (d). As the XC energy densities were 
extracted under the assumption of LDA the order of
magnitude of non-local effects is mapped in
Fig.~\ref{fig_unpolexample}(d). These problems have to be taken into
account if we interpret the results of extracted XC energy
densities. The best test for the reliability is the calculation of a
number of systems with different confinement potentials. In
Fig.~\ref{fig_unpol} we summarized the extracted $\varepsilon_{\rm
  XC}$ from systems with confinement potentials between $3\,{\rm
  meV}$ and $100\,{\rm meV}$. The $\varepsilon_{\rm XC}$ were taken
from the center and the density maximum. Thus, they are not
influenced by any edge effects and in addition they represent upper
and lower boundaries for the extracted XC energy densities (see
Fig.~\ref{fig_unpolexample} (d)). All curves show the difference between
a specific XC energy density and the TC XC energy density. While the
other parameterizations from literature (AMGB, ISI) do not deviate too
much from the TC reference both, the upper and lower bound of
extracted XC energy densities exhibit systematic corrections in the
regime of small density parameters or high densities. This trend is
stable and consistent for many different strengths of confinement
potentials and therefore it is reliable. The tendency to smaller XC
energy densities is also consistent with an increasing difference of
DFT and ED-GS energies with growing external confinement.  

\subsection{Fully polarized systems}

In the case of full polarization we consider the configurations with
$N=6$, $S=3$, $L=0$ and $N=8$, $S=4$, $L=0$ (see
Fig.~\ref{fig_motivation} for $N=8$). In analogy to the unpolarized
systems we invert the KS equations and solve the DEQ
(\ref{deq}). The resulting XC energy densities from the extrema
(center, maximum, minimum) are plotted in Fig.~\ref{fig_pol} (a) for six
electrons and in (b) for eight electrons. The strength of the
confinement potential varies between $3\,{\rm meV}$ and $100\,{\rm
  meV}$. The plots show again the
deviation from the TC parameterization. As in the case of unpolarized
systems we find for six and eight electrons systematic corrections
in the high density regime towards smaller XC energy densities. In
Fig.~\ref{fig_pol} (c) we check the consistency of the extracted XC
energy densities by comparing not only results for one system with different
confinement potentials but also different systems. The
good agreement is an important proof for the quality and the
reliability of our results.

\subsection{Partially polarized systems}

Finally we investigate the partially polarized systems $N=4$, $S=1$,
$L=0$ and $N=6$, $S=2$, $L=0$. As an example we consider four electrons
in a confinement potential of $3\,{\rm meV}$
(see Fig.~\ref{fig_partpolexample}). The spin-down and
spin-up Fig.~\ref{fig_partpolexample}(a) densities from ED are
exactly reproduced by the 
XC potentials plotted as dashed, black lines in
Fig.~\ref{fig_partpolexample} (b). Please note that the XC
potentials for the two spin-directions exhibit the same
asymptotic behavior as demanded. The grey, solid lines in (b)
are the (gauged) TC-XC potentials as they can be
calculated from the 
spin-densities in (a). After solving the DEQ (\ref{deq}) we
are able to plot the XC energy density as a function of the radius
(see (d)). However, for partial polarization we have to take into
account that $\varepsilon_{\rm XC}$ depends on both, $r_s$ and
$\xi$. 

In order to plot the results (Fig.~\ref{fig_partpol}) we have to
collect all combinations  
$(r_s,\varepsilon_{\rm XC})$ for given polarization $\xi$ from the
systems with four and six electrons and confinement energies from
$3\,{\rm meV}$ to $100\,{\rm meV}$. All the values were taken from areas with
a reasonably high density. Only in this regime results are reliable
as follows from the same arguments as for unpolarized and fully
polarized systems. The summary of extracted XC energy densities is
shown in Fig.~\ref{fig_partpol} for the polarizations $\xi=0.2$ (a),
$\xi=0.4$ (b), $\xi=0.6$ (c), and $\xi=0.8$ (d) as deviation from
the TC parameterization. Similar to unpolarized
and fully polarized systems we find systematic corrections for small
$r_s$ compared to TC, AMGB, and ISI. The reliability of the extracted
$\varepsilon_{\rm XC}$ is emphasized by the fact that the difference
between them is not larger than for $\xi=0$ and $\xi=1$. 

\section{Summary and outlook}
 
In this paper we present a new approach how to extract 2D-XC energy
densities from GS densities and energies of QDs calculated by ED. Our focus
was on configurations with arbitrary polarization but without
paramagnetic current density. In comparison with parameterizations
from literature (TC\cite{tanatar89}, AMGB\cite{attaccalite02},
ISI\cite{seidl01}) we find systematic corrections in the regime of
small density parameters or high densities for all
polarizations. This result was consistently accomplished by exploiting the
redundancy arising from different configurations and/or different
strengths of the external confinement potentials.

However, our results are restricted to a small range of density
parameters. Therefore, it would be useful to extend the calculations to
more confinement potentials or to confirm them by an alternative
method which also provides an approach for an improved
parameterization.  

Another open question is the role of the corrections due to the
paramagnetic current density or
vorticity\cite{vignale87, vignale88} which lead to an XC vector
potential. Up to now these effects have been considered in form of
complicated interpolations.\cite{levesque84, fano88, rasolt92} This problem
will be tackled in a forthcoming paper.

\section{Acknowledgement}
A.\ W.\ thanks the the RRZE Erlangen and the DFG (Ro 522/19-1).

\newpage

\noindent
\unitlength1cm
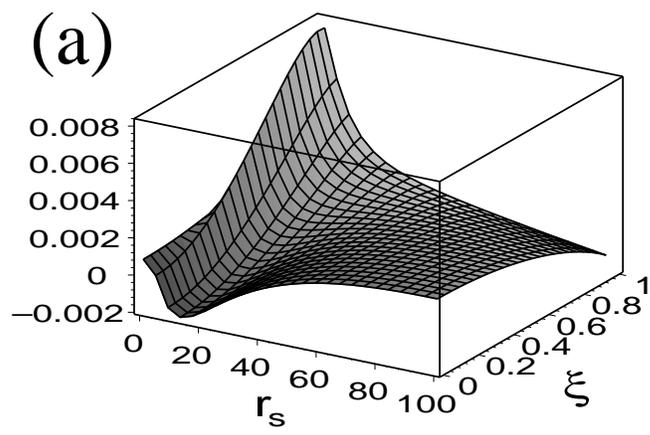
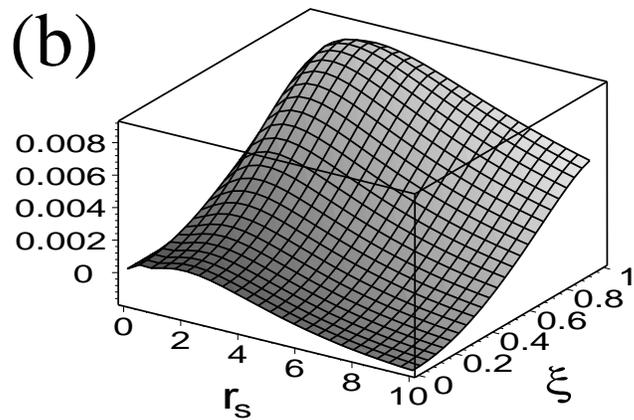
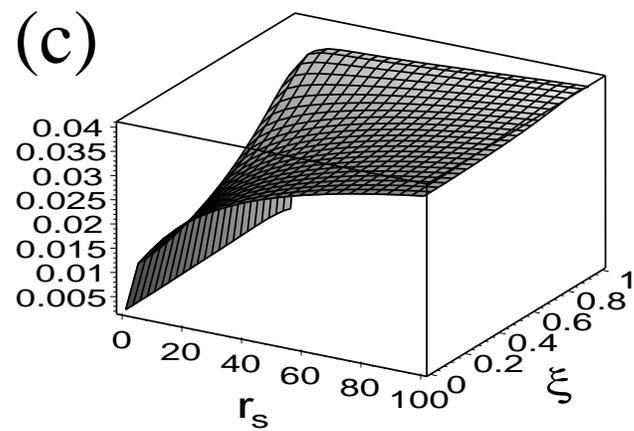
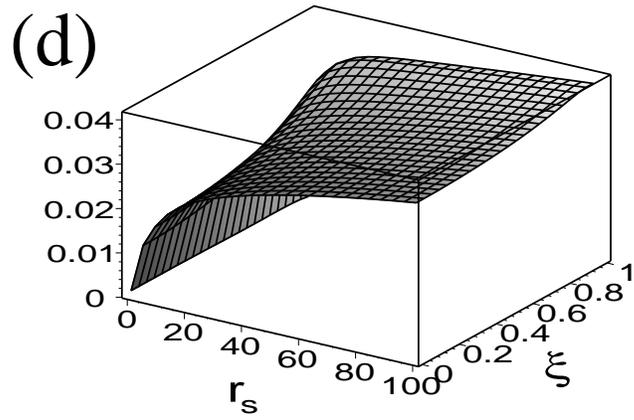
\begin{figure}
\caption{Comparison of TC\cite{tanatar89}, AMGB\cite{attaccalite02},
  and ISI\cite{seidl01} parameterizations as a function of density
  parameter and polarization. In (a) and (b) the
  relative deviation $\varepsilon_{\rm XC}^{\rm
  AMGB}/\varepsilon_{\rm XC}^{\rm TC}-1$ between TC and AMGB is
  plotted on different scales for $r_s$. (c) and (d) show the
  difference between ISI and TC ($\varepsilon_{\rm XC}^{\rm
  AMGB}/\varepsilon_{\rm XC}^{\rm TC}-1$) and ISI and AMGB
  ($\varepsilon_{\rm XC}^{\rm AMGB}/\varepsilon_{\rm XC}^{\rm TC}-1)$.
\label{fig_exccomp}
}
\end{figure}

\noindent
\unitlength1cm
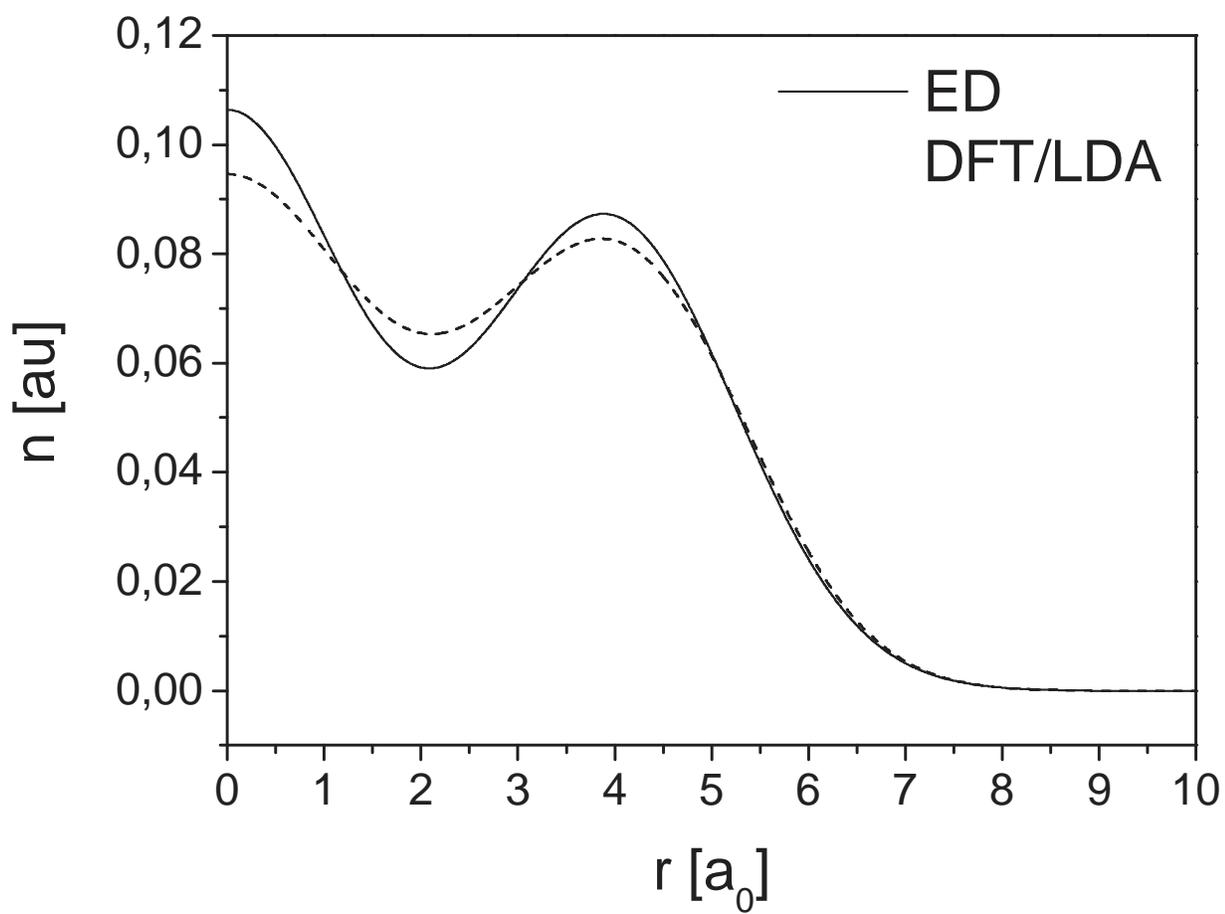
\begin{figure}
\caption{Comparison of GS densities from ED and DFT for a fully polarized
  eight-electron system ($L=0$, $S=4$) with pronounced differences
  in the density 
  extrema. The three DFT curves with TC, AMGB, and ISI
  parameterization lie so close that they cannot be resolved.
\label{fig_motivation}
}
\end{figure}


\noindent
\unitlength1cm
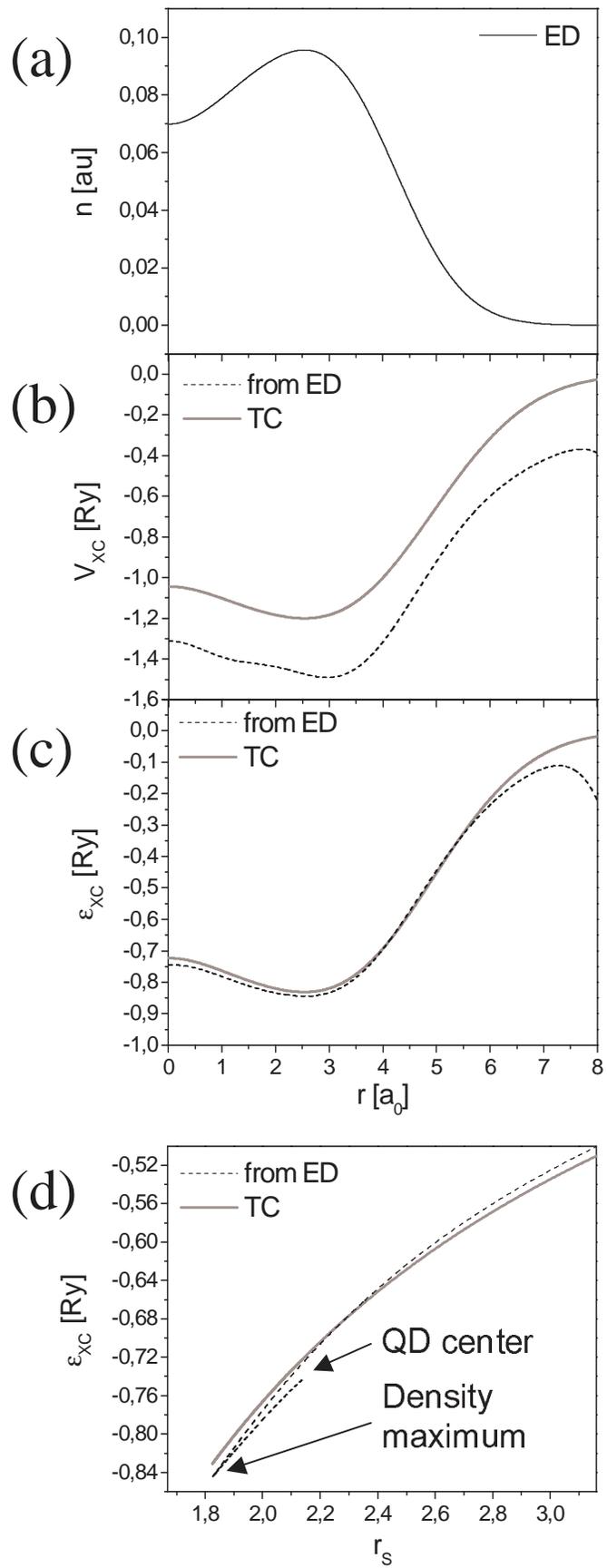
\begin{figure}
\caption{Numerical results of an unpolarized six-electron system with $L=S=0$
  and confinement potential $3.32\,{\rm meV}$. In (a) we show the GS density
  from ED in (a) as a function of the radius. It is reproduced by
  the XC potential (black, dashed line) in (b) which gives rise to
  the extracted XC 
  energy density  (black, dashed line) in (c). In (d) the extracted
  XC energy density is plotted 
  versus the density parameter $r_s$ (black, dashed line). The
  redundant structure is a 
  conseuquence of the density profile (see text). The grey, solid
  line in (b), (c), and (d) is the corresponding result calculated
  with TC parameterization. 
\label{fig_unpolexample}
}
\end{figure}

\noindent
\unitlength1cm
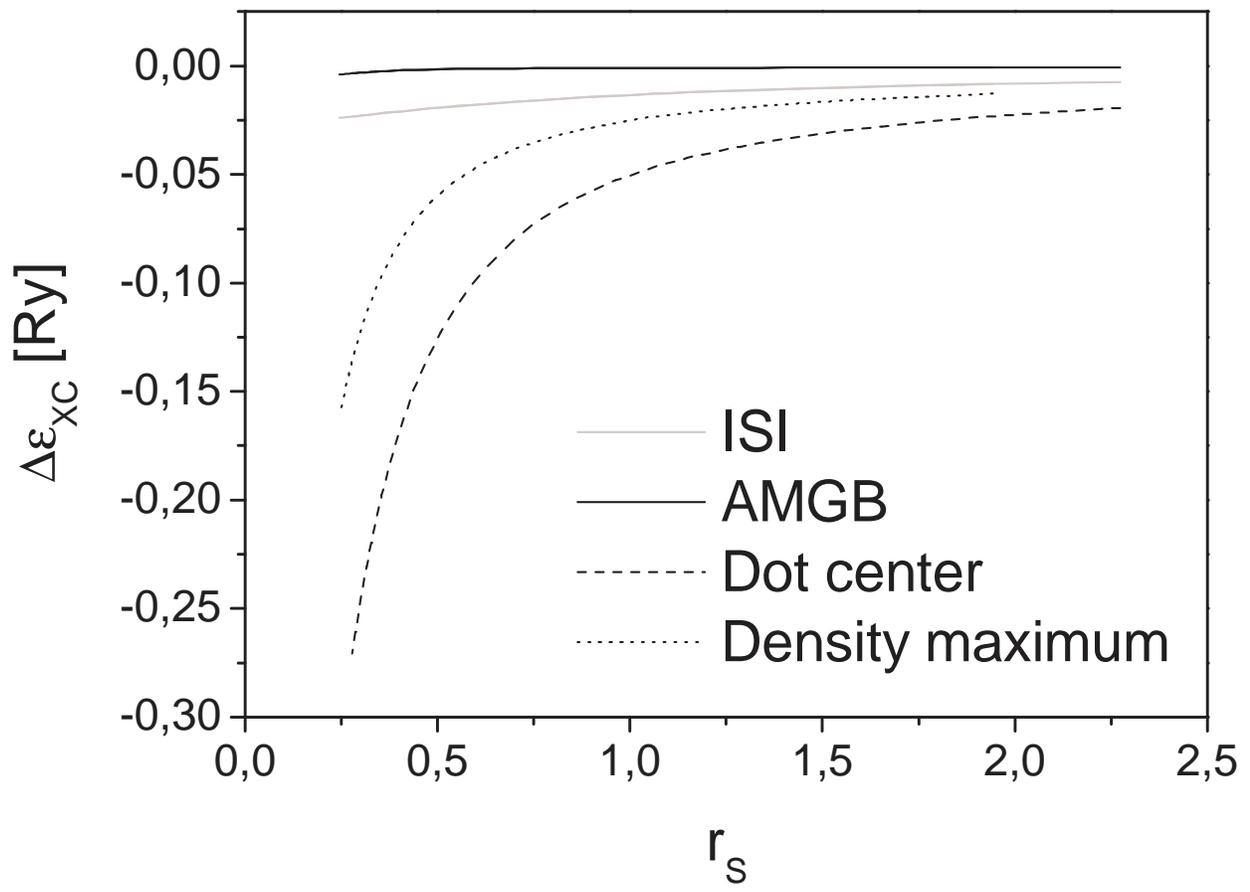
\begin{figure}
\caption{Summary of extracted XC energy densities for unpolarized
  systems (six electrons, different confinement potentials) as a
  function of $r_s$. We plotted the deviation of the 
  AMGB, the ISI, and the extracted XC energy densities from the TC
  parameterization. The results from the density maximum and the dot
  center are upper and lower bounds for the calculated
  $\varepsilon_{\rm XC}$-values. For small $r_s$ we find a
  systematic correction towards smaller XC energy densities.
\label{fig_unpol}
}
\end{figure}

\noindent
\unitlength1cm
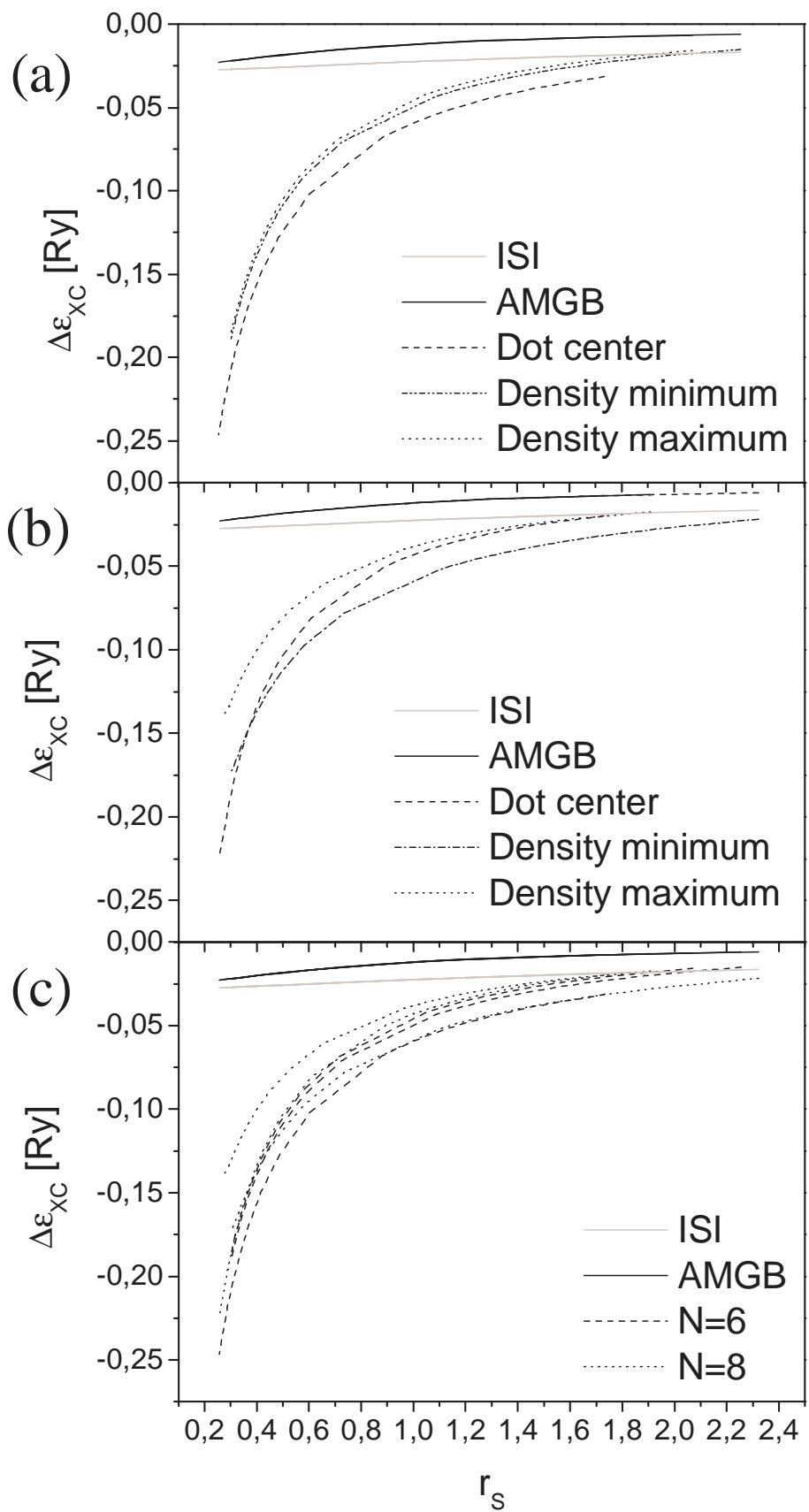
\begin{figure}
\caption{Summary of extracted XC energy densities for polarized
  systems (six (a) and eight (b) electrons, different confinement potentials)
  as a function of $r_s$. We plotted the deviation of the 
  AMGB, the ISI, and the extracted XC energy densities from the TC
  parameterization. The results from the density extrema represent
  upper and lower bounds for the calculated 
  $\varepsilon_{\rm XC}$-values. (c) shows
  the combined results for six and eight electrons and different
  confinement strength. For both
  electron numbers, we find consistent and systematic corrections towards
  smaller XC energy densities at small density parameters. 
\label{fig_pol}
}
\end{figure}

\noindent
\unitlength1cm
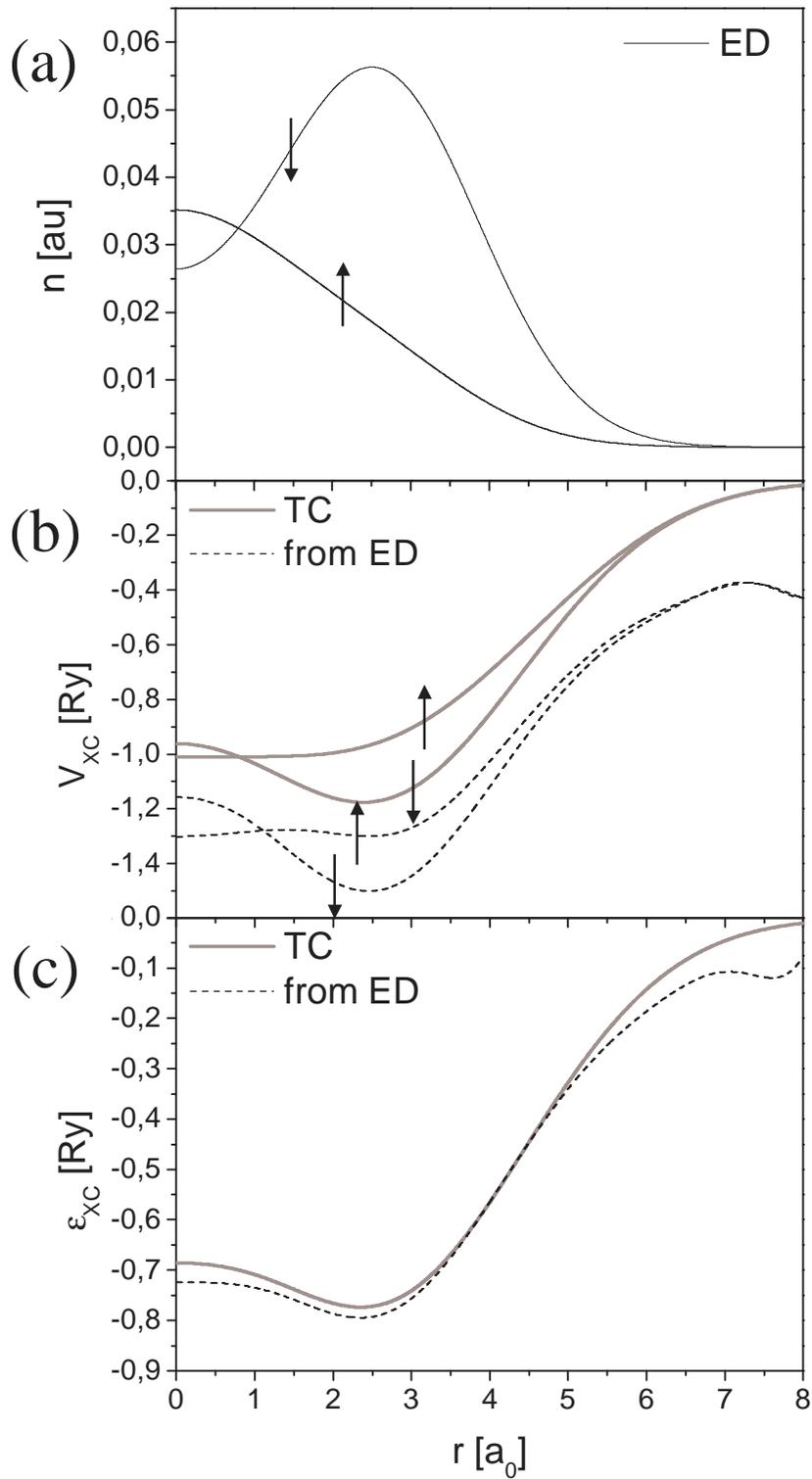
\begin{figure}
\caption{Numerical results of a partially polarized four-electron
  system with $L=0$, $S=1$ 
  and confinement potential $3\,{\rm meV}$. In (a) and (b) we show
  the GS densities 
  from ED for up- and down-spins as a function of the radius. They
  are reproduced by 
  the XC potentials (black, dashed lines) in (c) which give rise to
  the extracted XC 
  energy density  (black, dashed line) in (d). The grey, solid
  line in (b), (c), and (d) is the corresponding result calculated
  with TC parameterization. 
\label{fig_partpolexample}
}
\end{figure}

\noindent
\unitlength1cm
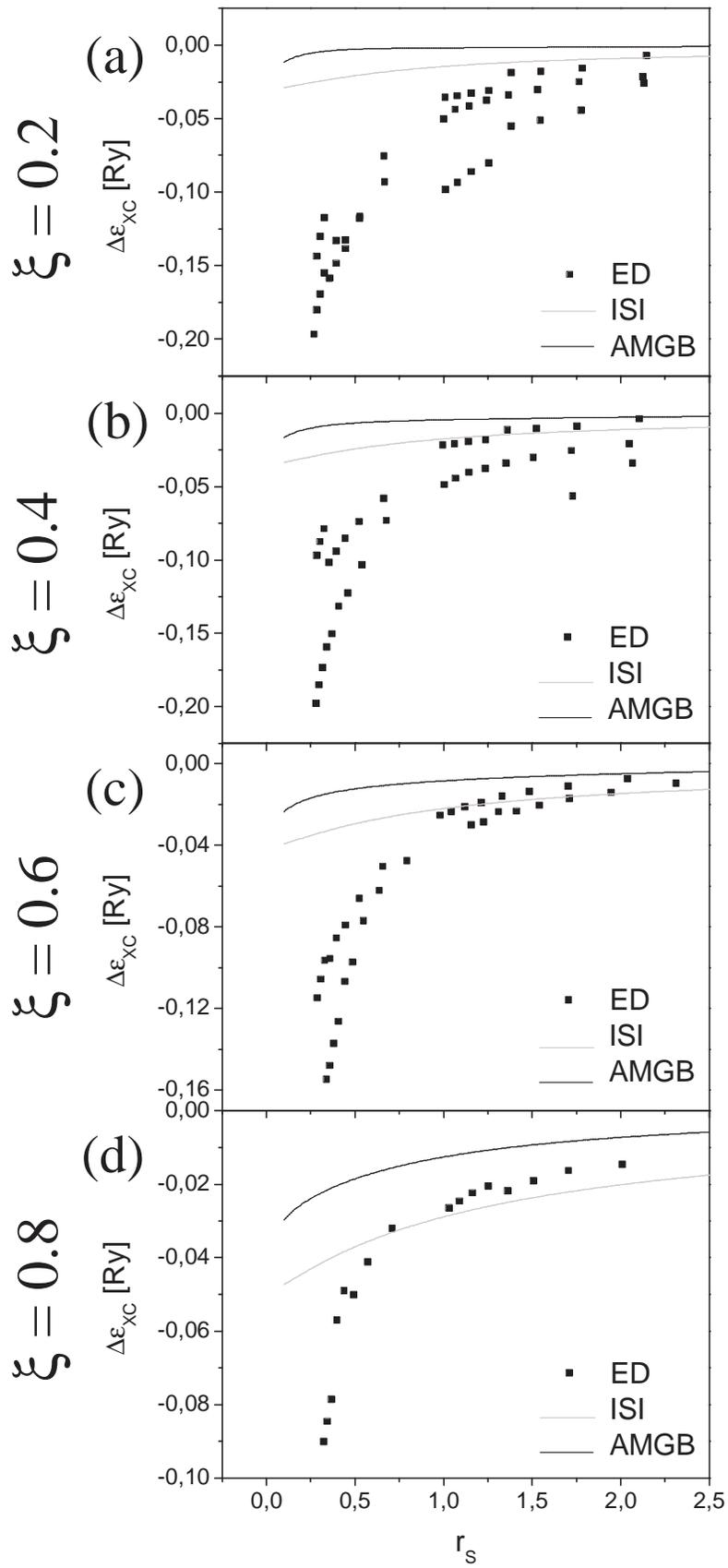
\begin{figure}
\caption{Summary of extracted XC energy densities for partially polarized
  systems (combined results for four and six electrons, different
  confinement potentials) 
  as a function of $r_s$ and polarization $\xi=0.2$ (a), 0.4 (b),
  0.6 (c) and 0.8 (d). We plotted the deviation of the 
  AMGB, the ISI, and the extracted XC energy densities from the TC
  parameterization. For all systems we find consistent and
  systematic corrections towards 
  smaller XC energy densities at small density parameters. 
\label{fig_partpol}
}
\end{figure}

\end{document}